\documentclass[10pt]{article}

\usepackage{amsmath}
\usepackage{amssymb}

\usepackage{graphicx}

\usepackage{cite}

\usepackage[labelfont=bf,labelsep=period,justification=raggedright]{caption}

\makeatletter
\renewcommand{\@biblabel}[1]{\quad#1.}
\makeatother

\date{}

\newcommand*{\R}{\mathbb{R}}

\usepackage{color}

\newcommand{\trackchange}[1]{#1}

\begin{document}

% Title must be 150 characters or less
\begin{flushleft}
{\Large
\textbf{Trade integration and trade imbalances in the European Union: a network perspective}
}
% Insert Author names, affiliations and corresponding author email.
\\
Gautier M. Krings$^{1,2}$, 
Jean-Fran\c cois Carpantier$^{3}$, 
Jean-Charles Delvenne$^{1,4,\ast}$
\\
\bf{1} Institute for Information and Communication Technologies, Electronics and Applied Mathematics (ICTEAM), Universit\'e{} catholique de Louvain, Avenue Georges Lema\^{i}tre, 4, 1348 Louvain-la-Neuve, Belgium
\\
\bf{2} Real Impact Analytics Belgium, Place Flagey 7, 1050 Brussels, Belgium
\\
\bf{3} CREA, University of Luxembourg, Campus Limpertsberg, BRC 0.02, 162 A, avenue de la Fa\"{i}encerie, L-1511 Luxembourg
\\
\bf{4} Center for Operation Research and Econometrics (CORE), Universit\'e{} catholique de Louvain, Voie du Roman Pays 34, 1348 Louvain-la-Neuve, Belgium
\\
$\ast$ E-mail: jean-charles.delvenne@uclouvain.be
\end{flushleft}

% Please keep the abstract between 250 and 300 words
\section*{Abstract}
We study the ever more integrated and ever more unbalanced trade relationships between European countries. To better capture the complexity of economic networks, we propose two global measures that assess the trade \trackchange{integration} and the trade imbalances of the European countries. These measures are the network \trackchange{(or indirect)} counterparts to traditional \trackchange{(or direct)} measures such as the trade-to-GDP (Gross Domestic Product) and trade deficit-to-GDP ratios. Our \trackchange{indirect} tools account for the European inter-country trade structure and follow (i) \trackchange{a decomposition of the global trade flow into elementary flows that highlight the long-range dependencies between exporting and importing economies} and (ii) the commute-time distance for trade \trackchange{integration}, \trackchange{which measures the impact of a perturbation in the economy of a country on another country, possibly through intermediate partners by domino effect}. Our application addresses the impact of the launch of the Euro. \trackchange{We find that the indirect imbalance measures better identify the countries ultimately bearing deficits and surpluses, by neutralizing the impact of trade transit countries}, \trackchange{such as the Netherlands}. \trackchange{Among others, we find that ultimate surpluses of Germany are quite concentrated in only three partners.} \trackchange{We also show that for some countries, the direct and indirect measures of trade integration diverge, thereby revealing that these countries (e.g. Greece and Portugal) trade to a smaller extent with countries considered as central in the European Union network}.
 
\section*{Introduction}
The structure of European trade has undergone deep transformations last two decades, such as the integration of former socialist economies and the introduction of the Euro. Across the years, under these events and the evolution of the world trade, the structure of European trade has been modified thoroughly, generally speaking in the sense of higher mutual trade but also in terms of acuter trade imbalance.

Among the indicators that can quantify the structure of trade, one finds direct measures of bilateral trade, such as the trade-to-GDP\footnote{Trade is the sum of imports and exports of goods and services. GDP denotes the gross domestic product.} ratio, which is a measure of \trackchange{the trade integration of} \trackchange{a country within its environment}, or such as the deficit-to-GDP\footnote{Trade deficit is the difference between imports and exports. A negative deficit is a trade surplus.} ratio, which is a measure of trade imbalance between \trackchange{a country and its partners}. These measures are \trackchange{direct} in the sense that they only take into account the interaction of one country with its trade partners, neglecting the remainder of the network. 

In this paper, we propose two complementary methods for analysing the global structure of a trade network. These methods take part in the development of complex networks theory for economics (see for example  \cite{Science, Soramaki2007} for general work in economics and \cite{HH2009,Saramaki2008,FRS2009,FRS2010,GL2005,HD2010} for specific works in trade). A key idea of complex networks theory is to acknowledge that the nodes of a network, such as the countries in a trade network, influence each other, not only directly (in a neighbour-to-neighbour relationship), but also indirectly (through intermediate nodes). Taking into account th\trackchange{e}se indirect influences may provide \trackchange{complementary insights to} the direct influences.

In the first method, we decompose the \trackchange{annual} flow of goods and services, quantified by their money counterpart measured in a common currency, into a sum of three coherent flows that illuminate different aspects of the trade network. The first flow is symmetric, which measures to which extent countries' trade are balanced on a pairwise basis. The second flow is cyclic, \trackchange{in which} every country's trade is globally balanced (but not on a pairwise basis). Such a flow may for instance exhibit a \trackchange{four}-way cycle where country $A$ exports to country $B$, while $B$ exports to country $C$, \trackchange{$C$ to $D$, and $D$ back to $A$} for the same amount. The third flow is acyclic, \trackchange{in which}  the money flow is a union of paths starting from a country in deficit and finishing to a country in surplus. If $A$ imports from $B$ that imports for the same quantity from $C$, we may consider that $C$ is the actual end creditor of $A$, rather than $B$ as the usual \trackchange{direct} measures of deficit would consider. The resulting \trackchange{indirect} deficits or surpluses between two countries provide a more comprehensive view of the \trackchange{global} trade between them than the mere \trackchange{direct} flow.

\trackchange{The relevance of considering the indirect flows can be illustrated with the example of the Netherlands. This country plays the role of a hub in the trade network, with imports from the rest of the world transiting through the Netherlands to the other European countries. The direct measure would only note that the Netherlands have a trade surplus with all European countries, neglecting that most of the trade only transits through the Netherlands from an initial exporter to a final importer. The indirect measure cleans these network-related effects to reveal the ultimate trade debtors and creditors. The indirect measure is thus also a relevant tool to assess the competitiveness of a country. This example illustrates the case of a chain of production with goods transiting through the Netherlands, but the relevance of the indirect measure is more general, and not confined to chains of production. The case of a country A exporting wheels to country B, which will then export cars (and the wheels herewith) to a country C is not fundamentally different from the case of a country A exporting beer to country B, letting him devote more resources (the ones previously affected to beer production) to wheel production and cars export to country C.} \trackchange{At the end, country A is the ultimate creditor and country C the ultimate debtor, given the resources the former used for the final consumption of the latter.} 

This Flow Decomposition Method and the associated notion of indirect trade deficit are the main conceptual innovations of this paper. \trackchange{These tools bear some superficial resemblance with the cycle, middleman, in and out statistics developed in \cite{Fagiolo2007}. The latter are defined from the direct interactions of three agents, e.g. the considered cycles are of length three ; it can therefore be seen as a bottom-up approach towards a network-theoretic view of trade, identifying patterns of trade just above the single node/edge level. By contrast our approach takes a top-down approach, explaining the total flow as a sum of smaller flows, yet possibly involving a large number of nodes, thus still global in nature.}

In the second method, we make a thought experiment in which we follow the random tribulations of a dollar travelling randomly through the trade network. At stationarity, the fraction of time spent by the random dollar, called the PageRank \cite{Page1999}, is a measure of the centrality of the country in the trade web. This measure of centrality is related to the GDP, which can be seen as a \trackchange{direct} measure of centrality, but corrects it by taking into account the centrality of the trade partners as well (the relevance of this measure in this context has already been evidenced \cite{RSF2008}). Taking the logic one step further, we consider the expected time for a dollar starting from country $A$ to reach $B$ and get back to $A$, which is called the commute time between $A$ and $B$, and can be seen as an economic distance between $A$ and $B$. More trade ties between $A$ and $B$, either direct or indirect (through common intermediate partners), correspond to a shorter distance, \trackchange{i.e. a tighter integration of these countries' economies}.  Using Principal Component Analysis---a classic spectral dimensionality reduction technique---we can visualize the countries as points in the plane, so that the usual \trackchange{square} Euclidean distance is approximately the commute-time distance. Repeating this figure year after year provides an animated picture of \trackchange{the world economy or a region thereof, where the evolution of individual or collective distances} can be tracked. In that the commute-time distance is symmetric, thus blind to the imbalances of particular trade relations, it acts as a complement to the first method. 

\trackchange{Trade integration between two countries is commonly apprehended through {bilateral} trade-to-GDP ratio.  The larger the ratio the more integrated the country with its partners. This traditional measure is direct in the sense that this statistics does not take into account the position, or centrality, of the trade partners in the network. 
Let us consider a mostly autarkic country $A$, trading only with country $B$. If $B$ trades with many partners, then $A$ is indirectly integrated within the trade network. In particular it gives $A$ access to the world markets, and makes it sensitive to the fluctuations of the global economy, much more than if $B$ were itself quasi-autarkic. The relevance of considering the indirect integration in the trade network will be illustrated in the applications where core and peripheral European countries follow distinct dynamics since the adoption of the euro. Indirect measures of the trade integration can be further interpreted as exposures to the propagation of a small perturbation of one country's activity to another country. This tool can be seen as one possible mathematical modelling of the shock propagation, through direct and domino effects, in a globalized economy.}

As a case study, we show how these \trackchange{indirect measures} complement their \trackchange{direct} counterparts in the context of the European trade. \trackchange{We especially study} the role of the Euro in the development of trade within Europe. The benefit of sharing a common currency (by decreasing the cost of cross-border transactions) depends on the degree of trade integration. The higher the trade \trackchange{integration}, the larger the benefits. It is therefore crucial to dispose of fine measures of trade integration. We show how our methods can better capture the complexity of the trade network as complement to the usual \trackchange{direct} measures.

%The article is organised as follows. In Section 2 we describe the network associated to European trade data. In Section 3 the Flow Decomposition Method for assessing global trade imbalance is presented and applied to the European trade network. The commute-time distance and associated visualisation, along with the economic interpretation and application to the European trade is explained in Section 4. Perspectives and conclusions are drawn in Section 5.

% You may title this section "Methods" or "Models". 
% "Models" is not a valid title for PLoS ONE authors. However, PLoS ONE
% authors may use "Analysis" 
\section*{Materials and Methods}
\subsection*{Data}
We study the network of trade flows between 24 countries of the European Union, using data provided by the International Monetary Fund (IMF) and the World Bank (WB). The two datasets provide different information about the network: the IMF dataset contains the total exports between each pair of countries of the European Union in current US Dollar value, while the WB dataset provides the gross domestic product (GDP) of each country and the consumer price index. By combining those two datasets, we build a network of trade flows between European countries. Since each export from a country $i$ to a country $j$ results in a flow of money in the opposite direction, we represent this data in a network of flows of currencies where each weighted and directed edge $(j,i)$ represents the total exports of good and services from $i$ to $j$. The internal demand of each country is represented as a loop $(i,i)$ given by GDP $-$(total exports $-$ total imports). All quantities are expressed in USD 2000, such that results are independent of inflation or exchange rate changes. Though expressing all data in a same currency is crucial, the choice of the currency (USD, EUR or other) is not relevant, given our methodology.

The data ranges from 1993 to 2007. By starting in 1993, we avoid the potential structural break related to the removal of EU internal customs (which had an impact on the recording of trade flows). By ending in 2007, we avoid the specific impact of the financial crisis. We therefore have a time period of 15 years.  Our dataset originally contains the data of the 27 members of the European Union. However, Malta and Cyprus have been removed from the dataset since large parts of their data were missing. Also, Belgium and Luxembourg are recorded together for the so-called Belgium-Luxembourg Economic Union (see Table \ref{tab1:Euro} for a list of the countries in the sample). 
Finally, one obtains a digraph $G(V,E)$ of 24 nodes ($V$) and 576 weighted directed edges ($E$) and loops represented by a weighted adjacency matrix $A:\{a_{ij}\}_{i,j=1}^{n}$, where $a_{ij}$ is the weight of the edge $(i,j)$. In this paper, we use the notation $G$ to represent the network and its matrix representation is written $A$.

\begin{table}[!ht]
\caption{
\bf{Countries sample}}
\centering
\begin{tabular}{lll}
\hline
\hline
	&Countries	&Category\\
\hline
1	 &Austria &	Euro\\
2 & Belgium-Luxembourg &	Euro\\
3	 &Finland &	Euro\\
4	 &France 	&Euro\\
5	 &Germany &	Euro\\
6	 &Ireland &	Euro\\
7	 &Italy 	&Euro\\
8	 &Netherlands& 	Euro\\
9&	 Portugal& 	Euro\\
10&	 Spain &	Euro\\
11&	 Greece 	&Euro\\
\hline
12&	 Slovenia& 	Non Euro\\
13&	 Slovakia &	Non Euro\\
14&	 Estonia 	&Non Euro\\
15&	 Bulgaria &	Non Euro\\
16&	 Czech Republic& 	Non Euro\\
17&	 Denmark 	&Non Euro\\
18&	 Hungary 	&Non Euro\\
19&	 Latvia 	&Non Euro\\
20&	 Lithuania &	Non Euro\\
21&	 Poland 	&Non Euro\\
22&	 Romania 	&Non Euro\\
23&	 Sweden 	&Non Euro\\
24&	 United Kingdom	&Non Euro\\
\hline 

\multicolumn{3}{p{7cm}} {\footnotesize{\textit{Notes}. Countries that did not adopt the Euro in 1999 (or 2002 for Greece) are classified as Non Euro countries. Euro and Non Euro countries are jointly referred to as European countries.}}
\end{tabular}
\label{tab1:Euro}
 \end{table}

\subsection*{Measuring \trackchange{indirect} trade imbalances with the Flow Decomposition Method}

Trade imbalance is usually apprehended through  trade deficit-to-GDP ratio. This measure is \trackchange{direct} in the sense that it only considers the situation with respect to direct trade partners. However, one could imagine countries having a trade deficit with countries that are themselves in deficit towards other countries, such that the deficits actually just transit through this country to a final trade creditor. We thus propose a global method emphasizing the trade imbalances from originary trade deficit countries to final trade surplus countries, therefore cleaning the picture from intermediate countries. This method relies on an original flow decomposition. 

\begin{figure}[!ht]
\begin{center}
 \includegraphics[width=10cm]{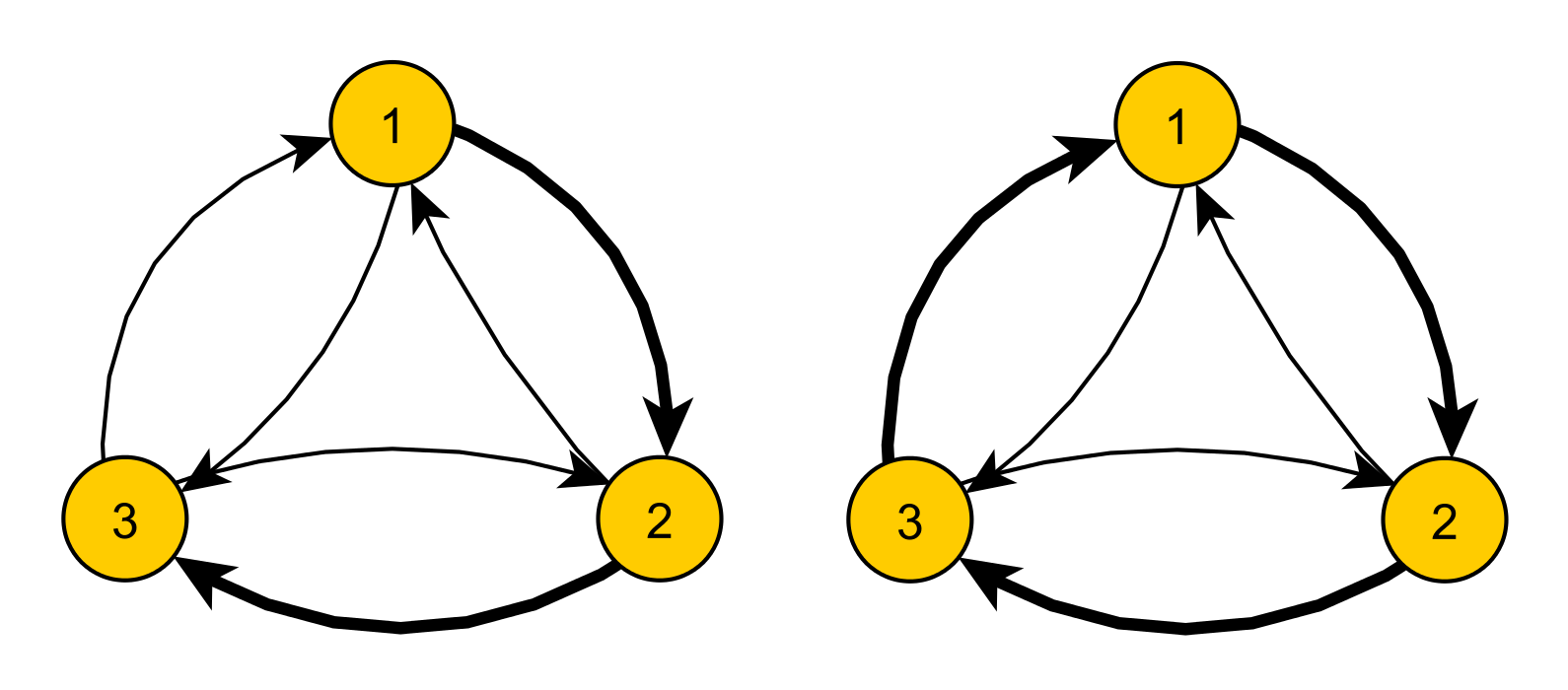}
\end{center}
\caption{{\bf Example of the influence of non-local flows.} We show two situations where node 1 has an apparent deficit towards node 2. On the left, the deficit of node 1 is transferred from node 2 to node 3, while on the right  the exchanges have been brought to an equilibrium.%and the interaction weights $a_{ij} = 2K$ and $a_{ji} =K$. The  comparison of those two weights suggest that there is a deficit of $K$ from $i$ to $j$. However, depending on the weights of $i$ and $j$ with $k$ the picture might change drastically: let us suppose that $a_{ik}=a_{ki}$ and $a_{jk}=a_{kj}+K$, the deficit of $i$ to $j$ is then transferred from $j$ to $k$, and thus the node $i$ has in fact an indirect deficit with respect to $k$. If, instead, $a_{ki} = a_{ik}+K$ while all other quantities remain identical, the deficit from $i$ to $k$ is transferred again to $i$, and in fact each country stays at balance.
}
\label{fig:example}
\end{figure}

We represent the trade network by its adjacency matrix $A:\{a_{ij}\}_{i,j=1}^{n}$; the entry $a_{ij}$ represents the amount of money flowing from country $i$ to country $j$, and the difference between $a_{ij}$ and $a_{ji}$ renders the imbalance in trade between both countries. However, this observation is only a first-order approximation of the exact asymmetry of trade, that does not take into account the relationship between the countries $i$ and $j$ and the rest of the graph. A very different picture might arise when looking at the graph as a whole. For example, let us consider the example in Fig. \ref{fig:example}, showing the interactions of three countries. In both examples, country 1 has a deficit towards country 2. However, depending on the value of the deficit of country 2 to country 3, the deficit of country 1 might be (or not) transferred to country 3, and hence country 3 is the actual creditor of country 1 rather than country 2. Finally, if country 3 has also a deficit towards country 1,  every country is in global balance even though every bilateral relationship is imbalanced and creates a cyclic circulation of value. \trackchange{Albeit very simple, this example} illustrates the need of extracting more complex information from the adjacency matrix than the net flow along single edges considered separately.

To do this, the trade network is separated in three components: symmetric, cyclic and acyclic, yielding the following expression for the adjacency matrices: $A=A^{S}+A^{C}+A^{AC}$.

These three components represent three kinds of interactions ongoing in the trade network: pairwise trade, cyclic trade and acyclic flows going from debtors to their \trackchange{indirect} creditors. We obtain these components from the original network through a succession of optimization models, that in succession provide us the symmetric ($G^S$), the cyclic ($G^C$) and the acyclic ($G^{AC}$) components. %We provide here only a sketch of these procedures, a detailed description of the optimization processes is provided in the Supplementary information.

The acyclic component is certainly the most insightful part of the trade decomposition. \trackchange{The  transfer of money from countries that accumulate deficits to countries that make surpluses is further decomposed into elementary flows from single country to single country that provide new insight on the exact transfers between individual countries. In particular any country is associated with a typically small number of final trade creditors or debtors, leading to a simpler, more readable picture than the one provided by direct deficits or surpluses, as illustrated in the Results and Discussion section.}

%(given by $y^{sd}$ in problem (\ref{eq:separate_flows}))

\subsubsection*{\trackchange{The symmetric flow}}

The symmetric component is obtained with \trackchange{the adjacency matrix $A^{S}$ whose entries are} $a_{ij}^{S}=\min (a_{ij},a_{ji})$. A simple observation shows that there is no larger network (larger in the sense that the weight of an edge could be larger than this solution) satisfying to the condition of symmetry. \trackchange{In other words, $a_{ij}^{S}$ it is the trivial, unique, solution $x_{ij}$ to the optimisation problem:}

\begin{equation}\label{eq:AS}
\begin{aligned}
& \underset{x_{ij}}{\text{maximize}}
& &\sum\limits_{ij}x_{ij} \\
& \text{subject to}
& & x_{ij} = x_{ji}, \; \forall i,j, \\
&&& 0 \leq x_{ij} \leq a_{ij}, \; \forall i,j.\\
%&&& x_{ij} \geq 0, \; \forall i,j.
\end{aligned}
\end{equation}

\subsubsection*{\trackchange{The cyclic flow}}

\trackchange{From the resultant network $G^{AS}$ of adjacency matrix $A^{AS}=A-A^S$---which does not contain any bidirectional edge---we extract the cyclic component by maximizing the flow circulating on the edges of $G^{AS}$ while imposing that the amount of incoming and outgoing flows in each node are equal, and that the flow on each edge does not exceed its weight. It can be written formally as a max-flow-like optimization problem}
\begin{equation}\label{eq:AC}
\begin{aligned}
z_1&&=& \underset{x_{ij}}{\text{maximize}}
& &\sum\limits_{i,j \in V}x_{ij} \\
&&& \text{subject to}
& & \sum\limits_{j \in V}x_{ij} - \sum\limits_{j \in V}x_{ji} = 0, \; \forall i, \\
&&& 0 \leq x_{ij} \leq a_{ij}^{AS}, \; \forall i,j \in V,\\
%&&& x_{ij} \geq 0, \; \forall i,j \in V.
\end{aligned}
\end{equation}
\trackchange{A} solution to this problem, found by standard linear programming methods, is \trackchange{a} cyclic component of the graph. \trackchange{However there may be typically many such cyclic flows, all with the same total value.} We want to choose a solution that is as spread out as possible among the different extreme solutions, as it is the most `neutral', not favouring one \trackchange{cycle} over another. A common `diversity index' \cite{hirschman1964paternity,simpson1949measurement} indicating how fairly a resource is distributed over several actors (e.g., market shares for companies or food for species) is given by the square of quantities attributed to every actor. For instance, if a unit resource is to be shared between three actors with quantities $\alpha$, $\beta$, $\gamma$, then the sum of squares $\alpha^2+ \beta^2 + \gamma^2$ is minimal when $\alpha = \beta = \gamma =1/3$. This is akin to a maximum entropy argument (even though the probabilistic context is not explicit here), as indeed the diversity index is equivalent to R\'{e}nyi's entropy \cite{renyi1961measures}. We therefore choose a particular maximum flow with the following sum-of-squares minimization problem:

\begin{equation}\label{eq:max_flow_cycl}
\begin{aligned}
z_2&&= & \quad \underset{x_{ij}}{\text{minimize}}
 & &\sum\limits_{i,j \in V}x_{ij}^2 \\
&&& \text{subject to}
& &\sum\limits_{i, j\in V}x_{ij} = z_1^{*}\\
&&& &&\sum\limits_{j\in V}x_{ij} - \sum\limits_{j\in V}x_{ji} = 0, \; \forall i \in V, \\
&&& & &0 \leq x_{ij} \leq a_{ij}, \; \forall i,j,\\
%&&& & &y_{ij} \geq 0, \; \forall i,j.
\end{aligned}
\end{equation}

 Here $z_1^{*}$ is the optimal cost of problem (\ref{eq:AC}). Simply stated, the solution of problem (\ref{eq:max_flow_cycl}) chooses among all optimal solutions of problem (\ref{eq:AC}) the cyclic flow that spreads out the most on the network.

\subsubsection*{\trackchange{The acyclic flow}}

The remaining links not taken into account in this solution form the acyclic network $G^{AC}$, of adjacency matrix $A^{AC}=A-A^{S}-A^{C}$.

\trackchange{The acyclic component can be further decomposed} into elementary flows, where each flow represents a certain amount of money flowing from a debtor country to a creditor country. These elementary flows represent how countries create debts or surpluses towards each other, not just directly but also through intermediates nodes. \trackchange{The economic relevance of indirect creditor/debtor relations, already discussed in the Introduction,  is illustrated below in the Results and Discussion section.}

The decomposition of the acyclic component into these elementary flows is typically non unique. 
We can always choose a decomposition such that every country is never \trackchange{both} a debtor to some countries and a creditor to some others. Indeed, if an elementary flow leaves from $i$ to $j$ and another flow from $j$ to $k$, then at least a part of this flow can be merged into a longer flow from $i$ to $k$.
\trackchange{Such a decomposition has the advantage of simplicity of interpretation, as every country is either a pure debtor or a pure creditor, or none.}

%\trackchange{[JCD: the next paragraphs are adapted from the Supp Inf]}

 Among all possible such decompositions, we choose the one that spreads the flow between two countries as much as possible on all possible trade routes, \trackchange{similarly to the cyclic decomposition}. We now give the details of how the elementary flows are computed. 
 
We start from an arbitrary acyclic weighted graph $G$ (as we suppose that we have already removed the symmetric and cyclic parts), consisting of a set of nodes $V$ and a set of edges $E$ and represented by its weighted adjacency matrix $A=\{a_{ij}\}_{i,j=1}^{|V|}$. \trackchange{The algorithm proceeds by repeated decompositions into single-source single-destination flows, which we call elementary flows. In this context, sources are debtors and destination are creditors.} 

\trackchange{In the first iteration, the sources $S$ are taken as the zero-indegree nodes and the destinations $D$ are the zero-outdegree nodes. Without loss of generality, we assume that the flow is connected (otherwise we treat connected components one after another). This ensures in particular that no node is  both a source and a destination.  Elementary flows are computed between $S$ and $D$, in the way that we explain hereafter. Then those flows are withdrawn, revealing a smaller flow to be decomposed, with possibly new sets of sources and destinations, on which a new iteration is performed, and so on until the entire acyclic flow has been decomposed.} One single iteration of this algorithm consists of three steps, which we now detail.

 First, the largest possible flow from the sources to the destinations is computed with a classic maximum flow problem:

\begin{equation}\label{eq:max_flow}
\begin{aligned}
w_1&&= & \quad \underset{x_{ij}}{\text{maximize}}
 & &\sum\limits_{i \in S,j\in V}x_{sj} \\
&&& \text{subject to}
& &\sum\limits_{j \in V}x_{ij} - \sum\limits_{j \in V}x_{ji} = 0, \; \forall i \in V \backslash \{D \cup S\}, \\
&&& && 0 \leq x_{ij} \leq a_{ij}, \; \forall i,j \in V\\
%&&& && x_{ij} \geq 0, \; \forall i,j.
\end{aligned}
\end{equation}

A solution to this problem is not unique in general. It is possible, for instance if there exists a bottleneck in the network, that the max flow goes from one source to one destination, or from another source to another destination, or may be shared between the two paths.  Similarly to the cycle decomposition, we therefore choose a maximum flow that is as spread out as possible among the different extreme solutions, with the following sum-of-squares minimization problem:

\begin{equation}\label{eq:max_flow_quad}
\begin{aligned}
w_2&&= & \quad \underset{y^{sd}, x^{sd}_{ij}}{\text{minimize}}
  &&\sum\limits_{s \in S,d\in D}(y^{sd})^2  \\
&&& \text{subject to}
&  & \sum\limits_{s \in S, d\in D}y^{sd} = w_1^{*}, \\
&&& &&\sum\limits_{j \in V}x^{sd}_{sj} = y^{sd}  \; \forall s \in S, d \in D,\\
&&& &&\sum\limits_{j\in V}x^{sd}_{ij} - \sum\limits_{j\in V}x^{sd}_{ji} = 0, \; \forall i \in V \backslash \{S \cup D\}, s \in S, d\in D, \\
&&& &&\sum\limits_{s\in S, d\in D}x^{sd}_{ij} \leq a_{ij},\; \forall i,j\in V,\\
&&& &&0 \leq x^{sd}_{ij}, \; \forall i,j \in V ,s\in S,d\in D.\\
\end{aligned}
\end{equation}

 Here $w_1^{*}$ is the optimal cost of problem (\ref{eq:max_flow}). The solution of problem (\ref{eq:max_flow_quad}) decomposes the total flow from sources to destinations into individual flows from single source $s$ to single destination $d$, of value $y^{sd}$, and chooses to spread the maximum flow $w_1^{*}$ as much as possible between the individual single-source single-destination flows $y^{sd}$. 
 
 \trackchange{We want further determine the intermediate nodes and edges by which the flow $y^{sd}$ circulated, i.e. to determine the contribution  $x^{sd}_{ij}$ on edge $ij$ to the flow $y^{sd}$. As once again those contributions can be chosen in several ways, we find the most spread out decomposition by minimizing the sum of squares of $x^{sd}_{ij}$ while maintaining the previously found optimal values $y^{sd*}$ for the single flows.}
  
\begin{equation}\label{eq:separate_flows}
\begin{aligned}
w_3&&= & \quad \underset{x^{sd}_{ij}, x_{ij}}{\text{minimize}}
 & &\sum\limits_{s \in S,d\in D}(x^{sd}_{ij})^2  \\
&&& \text{subject to}
& &\sum\limits_{j \in V}x^{sd}_{sj} = y^{sd*}  \; \forall s\in S,d\in D,\\
&&& &&\sum\limits_{j\in V}x^{sd}_{ij} - \sum\limits_{j\in V}x^{sd}_{ji} = 0, \; \forall i \in V \backslash \{S \cup D\}, s \in S, d\in D, \\
&&& &&\sum\limits_{s\in S, d\in D}x^{sd}_{ij} \leq a_{ij},\; \forall i,j\in V,\\
&&& &&0 \leq x^{sd}_{ij}, \; \forall i,j \in V ,s\in S,d\in D.\\
\end{aligned}
\end{equation}

This ends the iteration. By subtracting the flows on each edge (given by \trackchange{$x_{ij}^{sd}$}) from the original graph, one results in several disconnected components, on which the same process can be applied recursively until the graph has been totally decomposed.

\subsubsection*{\trackchange{Computation and example}}

\begin{figure}[!ht]
\begin{center}
 \includegraphics[width=10cm]{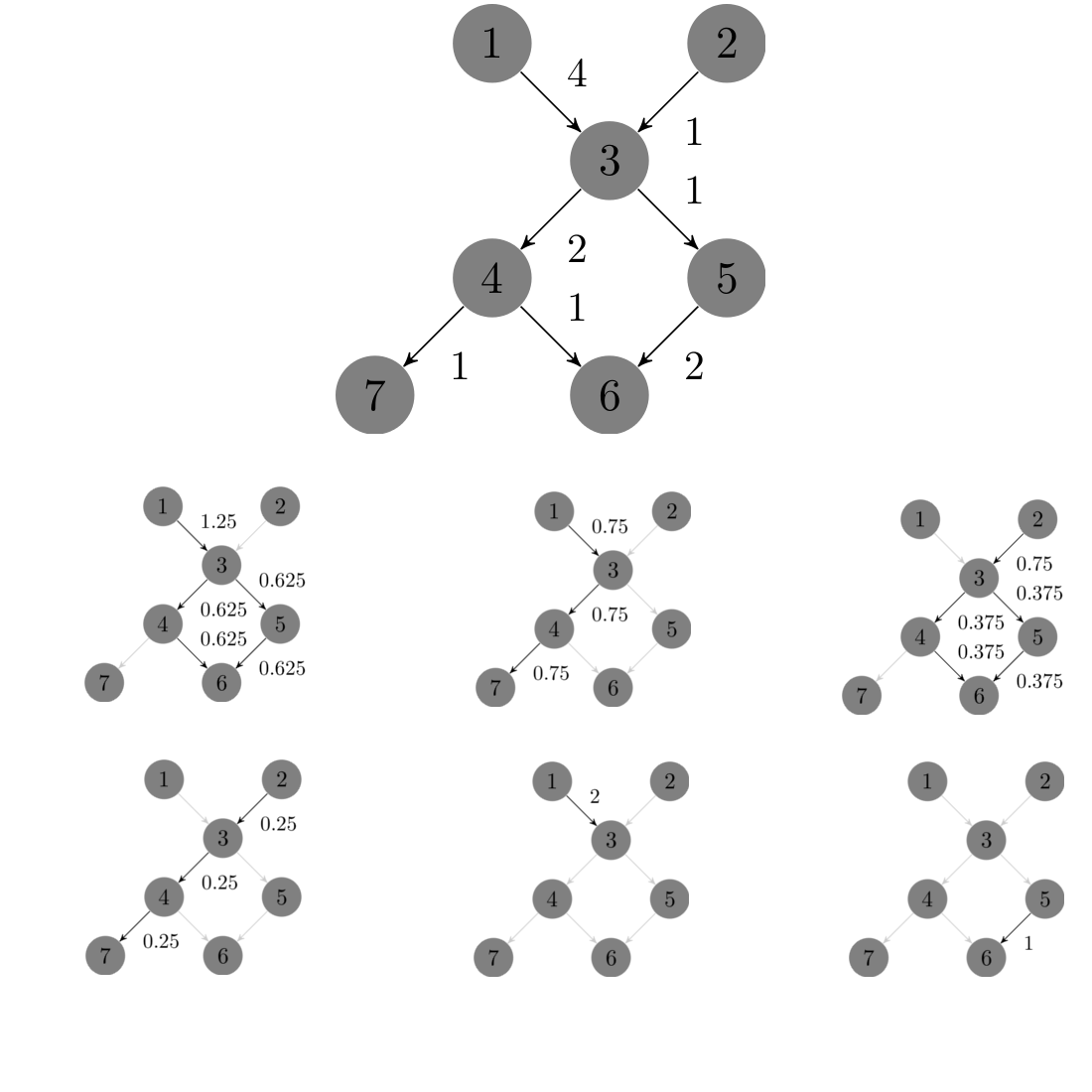}
\end{center}
\caption{Example of a decomposition in flows. The original graph (top) is acyclic, and is decomposed in six flows, such that the flows are most spread out, by a repeated application of problems (\ref{eq:max_flow}) to (\ref{eq:separate_flows}).}  
\label{fig:decompositionex}
\end{figure}

Algorithmically, the optimisation problems (\ref{eq:AC}) to (\ref{eq:separate_flows})  take the form of linear or convex quadratic programs that can be solved efficiently with usual methods.  An example of such a decomposition is given on Fig. \ref{fig:decompositionex}, where a flow on the graph is decomposed into 6 different flows, from 3 different sources to 3 different destinations.

\subsection*{Measuring \trackchange{indirect} trade \trackchange{integration} with the Commute Time Distance Method}

The trade \trackchange{integration} \trackchange{of a country with another country or group of countries} is usually measured by the bilateral trade-to-GDP ratio. The larger the ratio the more \trackchange{integrated} the country \trackchange{with its partners}. This measure is \trackchange{direct} in the sense that this statistics does not take into account the position, or centrality, of the trade partners in the network. We propose to use commute \trackchange{times} of a random walking dollar in the trade network as a measure of \trackchange{integration of countries with respect to one another taking into account their direct and indirect ties.} \trackchange{A small commute time between two countries implies a small trade distance between those countries, i.e. a high mutual integration.}

\subsubsection*{The random dollar and commute-time distances}

The commute-time distance, defined in the general context of random walks over graphs or Markov chains \cite{grinsteadintroduction, boley2011commute}, proceeds in the present case from a thought experiment. 

We imagine that all the money  in circulation in the world is materialised in one-dollar notes. We identify one particular note and we track it as it switches from hand to hand. We suppose that every dollar paid out by an individual is randomly uniformly chosen among all the dollars in her possession.  For the moment we suppose also time invariance: every two individual who have economic contacts, have regular economic contacts whose frequency and intensity (measured for instance by the number of dollars exchanged in a year) is constant in time, thus resulting in a constant expected fortune (number of dollars possessed) of individuals through time. The probability of presence of a random walker in a given node \trackchange{of} a network is sometimes called the PageRank of the node, in analogy with a well known search tool on the World Wide Web \cite{Page1999}, and gives a measure of its centrality in the network. In this case the PageRank  is directly proportional to the expected fortune of an individual.

 In our case, nodes represent individual countries, and we assume that the random dollar is exchanged at regular discrete-time steps, rather than in continuous time. \trackchange{We can choose a restricted network (e.g., Europe) and perform a  random walk on restricted network, which corresponds to the random walk in the global network conditional to the fact that the random dollar stays in the restricted network.}
In this paper, we study the random dollar walk conditional to the fact that it corresponds to a trade of goods and services \trackchange{produced in the year} within Europe. This excludes the non-European countries but also \trackchange{the trade of second-hand goods already accounted for in previous years}.

%Note that we consider the part of GDP of a country $A$ that is neither exported nor imported as being `exported from $A$ to $A$', therefore every country has a loop of intensity GDP $-$ (total exports $-$ total imports). 

In this network, the PageRank is a truly global quantity which gives a useful indication of the centrality of a country in the network of exchanges, as exemplified in \cite{RSF2008}. In this section we enrol another classic tool of random walk theory, namely commute \trackchange{times}, at the service of trade networks. 

Let $\tau_{AB}$ be the so-called hitting time from country $A$ to country $B$, which is the time of first passage to $B$ for a random note that would start in $A$. The commute-time distance between countries $A$ and $B$ is defined as the square root of the commute time $\tau_{AB}+\tau_{BA}$. It is a Euclidean distance, which means that every of the $n$ countries can be embedded in $\R^n$, so that the usual  distance between the two points representing $A$ and $B$ coincides with the commute-time distance.

\subsubsection*{\trackchange{Interpretation of commute-time distances}}

\trackchange{The Euclidean commute-time distances embeds the national economies as points in a geometric space, where proximity of two countries indicates that a large fraction of the GDP of both countries participes to trade between them, either directly or through a chain of intermediate partners.} \trackchange{Assuming that $A$ trades actively with $B$, which interacts actively with $C$ (or partners of $C$), then the commute-time distance between $A$ and  $C$ can be relatively short, indicating fairly good mutual integration, even though the direct trade-to-GDP ratio of $A$ towards $C$ is zero.}

\trackchange{Commute times can be further interpreted in terms of the propagation of a small perturbation of one country's activity to another country, as the following simplistic thought experiment suggests. Imagine indeed that a new activity is created in country $A$, and a new dollar is printed that reflects the value of this new activity. In the absence of any knowledge of the nature of this activity, we may suppose that this new dollar will diffuse randomly in the network. The hitting time therefore indicates the speed at which the infinitesimal burst of activity in $A$ is felt in another country $B$. The same argument goes for an infinitesimal destruction of activity. Therefore our measure of mutual integration of two countries' economies may be identified with the mutual sensitivity of each country to the other's economic health.}

\subsubsection*{\trackchange{Integration within a group of countries}}

\trackchange{So far we have presented commute-time distance as a pairwise relationship indicating mutual integration of two countries. However we are often more interested in assessing how a country is integrated in the regional or global economy, rather than with a specific partner. Clearly, a regional economy  is tightly integrated as a whole, if all the national economies are close to each other in terms of commute-time distances,} \trackchange{which translates into a quick propagation of shocks across the network through direct or domino effects.} \trackchange{This can be assessed by the variance of the countries in their Euclidean space representation, which is the average square commute-time distance to the centre of mass of the region. The square distance of one particular country to the centre of mass indicates how integrated this country is with the regional network as a whole, rather than with a specific country.}

\subsubsection*{\trackchange{Computation and representation}}

\trackchange{We now explain how to compute the commute-time distances}. Let $Imp$ be the import matrix, where ${Imp}_{AB}$ is the amount of imports of $A$ from $B$, corresponding to a money flow from $A$ to $B$. The diagonal ${Imp}_{AA}$ is $GDP_A + Imports_A - Exports_A$, \trackchange{following the idea that the value produced in $A$ but not imported nor exported is akin to an export from $A$ to $A$}. The random dollar's probability transition matrix $P=D^{-1}{Imp}_{AB}$, where $D$ is the diagonal matrix of strengths: $D_{AA}$ is the sum of row $A$'s entries of $Imp$. Let us illustrate this formula on a simplistic example. Consider a country $A$ importing steel from country $B$ for 70 USD and making for 100 USD worth of cars, of which 60 USD are exported and 40 USD are bought locally. The GDP is $100-70=30$ USD, and ${Imp}_{AA}=40$. Then the random dollar will jump from $A$ to $B$ (as counterpart to steel import) with probability $70/110$ and will stay in $A$ (as a counterpart for a car sale) with probability $40/110$. The PageRank vector is given by $P$'s dominant left eigenvector $\pi$, obeying $\pi=\pi P$. If we call $\Pi$ the diagonal matrix constructed from $\pi$, then the Laplacian \cite{boley2011commute} is $\Pi - \Pi P$, and the commute time between $A$ and $B$ is $\tau_{AB}+\tau_{BA}=(e_A-e_B)^T L^+ (e_A-e_B)$, where $e_A$ is the column characteristic vector of country $A$ (with zeroes everywhere excepted a one at entry $A$) and $L^+$ is the pseudo-inverse of the Laplacian \cite{boley2011commute}. Evidently, the matrix $L^+$ can be replaced by its symmetric part $(L^+ + {L^+}^T)/2$. 

 In order to facilitate the visualisation or analysis of data, one can perform a principal component analysis of the $n$ points into $k$ dimensions (for any $k<n$), which amounts to replacing $(L^+ + {L^+}^T)/2$ by its rank $k$ approximation, thus placing the $n$ nodes in a $k$-dimensional plane. Usually $k=2$ is taken for visualisation purposes, if relevant. This provides an economic interpretation of a classical method of representation of a network \cite{hall1970r} called Eigenprojection method in \cite{Koren2005}. Repeating the procedure for the data of each year results in a series of images that picture the evolution of European economies \trackchange{year after year}.

% Results and Discussion can be combined.
\section*{Results and Discussion}

\subsection*{\trackchange{Indirect} trade imbalance in Europe}

We now illustrate how the tools derived from the flow decomposition approach can help to better understand the European trade network. 
The symmetric, cyclic and acyclic components of the inter-country trade structures represent three types of economic interactions. The symmetric part reflects the pairwise exchanges where both countries have neither trade deficit nor surplus. The cyclic part reflects the same kind of interaction, but circulating over a larger number of actors\trackchange{. S}ymmetric interactions can be seen as cycles of length 2. Finally, the third part contains the transfers of money that flow from countries that accumulate deficits to surplus-making countries. The study of those three components provides insights on the evolution of the European trade network.

\begin{figure}[!ht]
\begin{center}
 \includegraphics[width=\textwidth]{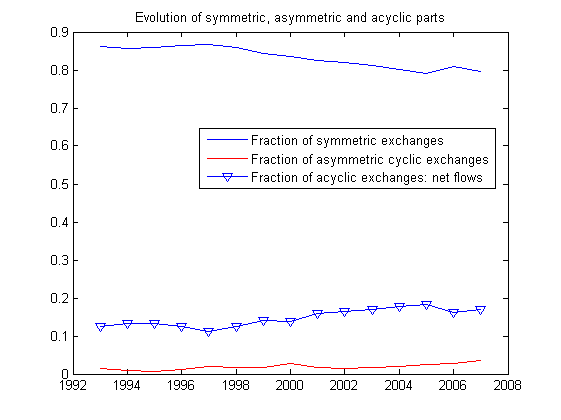}
\end{center}
\caption{{\bf Flow decomposition of trade in the Eurozone}. \trackchange{Decomposition of the annual trade flows into symmetric, asymmetric cyclic and asymmetric acyclic components.}}
\label{fig:flow}
\end{figure}

Firstly, the largest fraction of the European trade is made of symmetric exchanges (Fig. \ref{fig:flow}), which reflects the tendency of countries to restrain, to a certain extent, their imbalances. \trackchange{A fraction of, say, 90\% of symmetric exchanges means that, on average, bilateral direct imbalances are equal to 10\% of total trade}. Secondly, as regards the fraction of cyclic exchanges, its part remains close to zero, indicating that balanced clusters of countries are just a small share of the total trade network. The trend in the very last years is slightly rising, which indicates a growing tendency of countries to trade and balance their trades inside cluster of countries, rather than on bilateral basis. Finally, the fraction of acyclic exchanges increases significantly from 1997 onward.

\begin{figure}[!h]
\begin{center}
 \includegraphics[angle=90,width=13cm]{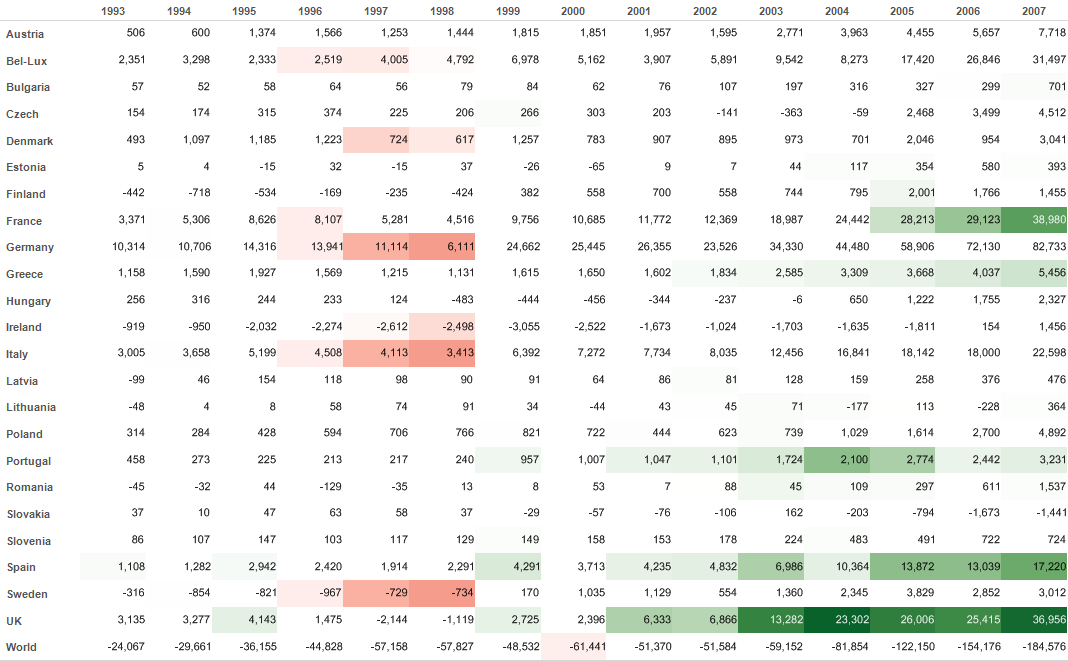}
\end{center}
\caption{ \trackchange{{\bf Evolution of the direct and indirect measures of trade imbalances for the Netherlands.} The figures in each cell correspond to \trackchange{direct} trade surpluses (+) or deficits (-) of the Netherlands toward countries listed on the rows. The colors correspond to the indirect measures of trade imbalances, as computed by the Flow Decomposition Method, with ultimate surpluses in green and ultimate deficits in red.}}
\label{colorsandnumbers_NED}
\end{figure}

This acyclic component provides a new way to consider the trade imbalances in a network of countries.  \trackchange{Contrarily to the traditional approach focusing on the trade deficit-to-GDP between two countries, a direct measure, the tool we propose goes farther by considering} \trackchange{the indirect exposures to get at the end a finer understanding of the ultimate debtors and creditors of trade flows.} \trackchange{The example of the Netherlands,} \trackchange{see Fig. \ref{colorsandnumbers_NED},} \trackchange{is illustrative in this respect. Considering the direct measures (trade deficit-to-GDP), the Netherlands have a \trackchange{growing} trade surplus toward Germany and Spain over the whole period. These trade surpluses can reflect a rising relative competitiveness, with a trade surplus emerging from rising exports and falling imports, but they can also reflect network effects not specific to these pairs of countries. Indeed, trade surpluses can also be related to the general organization of the world trade and the respective specialization of the countries in the world production.} \trackchange{The Netherlands for example play the role of a hub in the European trade by importing huge volumes from the world and redistributing these in Europe through exports. The indirect debtor/creditor relationships we propose} \trackchange{precisely accounts for the network effects in view to find the ultimate creditors of Spain or Germany that trade with them through the Netherlands, and cleans the relationships between the Netherlands and Spain or Germany from those flows.} 
%\trackchange{Figure \ref{colorsandnumbers_NED} illustrates this case by combining the direct and indirect measures, with deficit values in numbers for the former and colours for the later. 
\trackchange{Though trade surpluses of the Netherlands towards Spain and Germany persist over the whole period according to the direct measures, an indirect surplus is only reported toward Spain.}

\begin{figure}[!h]
\begin{center}
 \includegraphics[angle=90,width=13cm]{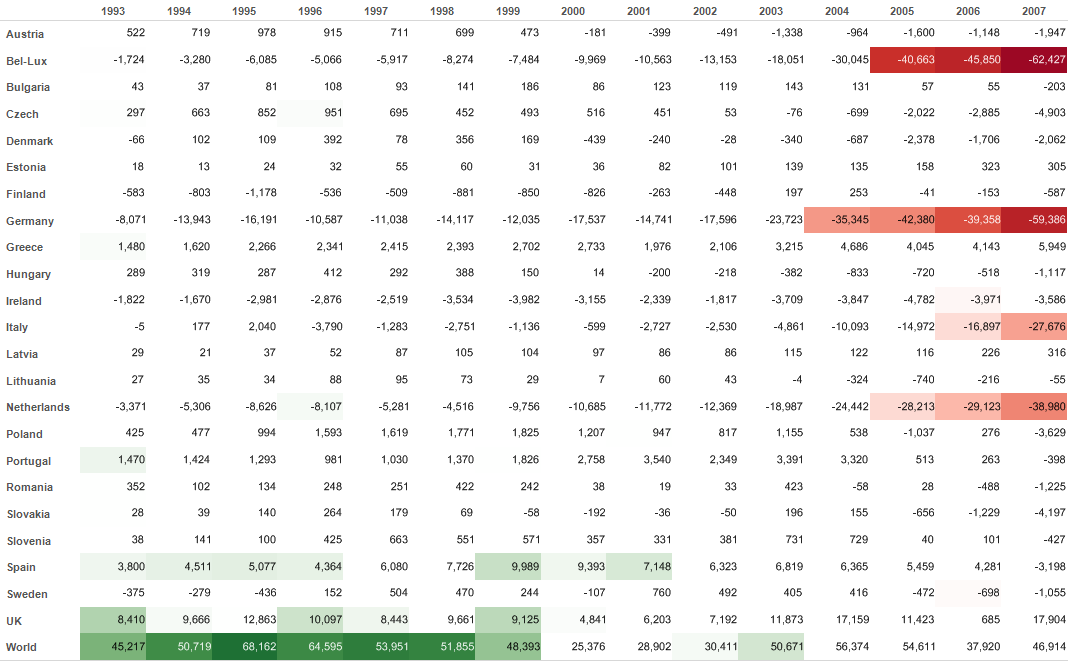}
\end{center}
\caption{ \trackchange{{\bf Evolution of the direct and indirect measures of trade imbalances for France.} The figures in each cell correspond to \trackchange{direct} trade surpluses (+) or deficits (-) of France toward countries listed on the rows. The colors correspond to the indirect measures of trade imbalances, as computed by the Flow Decomposition Method, with ultimate surpluses in green and ultimate deficits in red.}}
\label{colorsandnumbers_FRA}
\end{figure}

\trackchange{The relevance of the indirect measure of the trade deficit is double. First, it reveals ultimate deficits/surpluses, cleaned from} \trackchange{transit node effects}. As such, the indirect measure is an analytic complementary tool to the direct trade deficit-to-GDP measure. The indirect measure should better capture specific competitivity problems since it controls for country specializations in the trade network\footnote{The bilateral \trackchange{direct} trade deficit-to-GDP is a noisy measures to the extent that a deficit can reveal both a relative competitiveness problem and a specific role in the world trade network (a specialization or a role of hub). The indirect measure by neutralizing the network component of the deficit can thus be considered as a more precise measure of competitiveness }. Secondly, it reduces the size of the information matrix. The indirect measures reduce the bilateral trade surpluses/deficits matrix to a \trackchange{relatively small} \trackchange{subset of ultimate bilateral trade deficits/surpluses. Countries who measure the evolution of their competitiveness can thus focus on the sole countries with which indirect trade deficits prevail.} \trackchange{For example, Fig. \ref{colorsandnumbers_FRA} shows that France may focus on restoring competitiveness with respect to Germany and Belgium-Luxembourg in priority.} 

\begin{figure}[!h]
\begin{center}
 \includegraphics[angle=90,width=13cm]{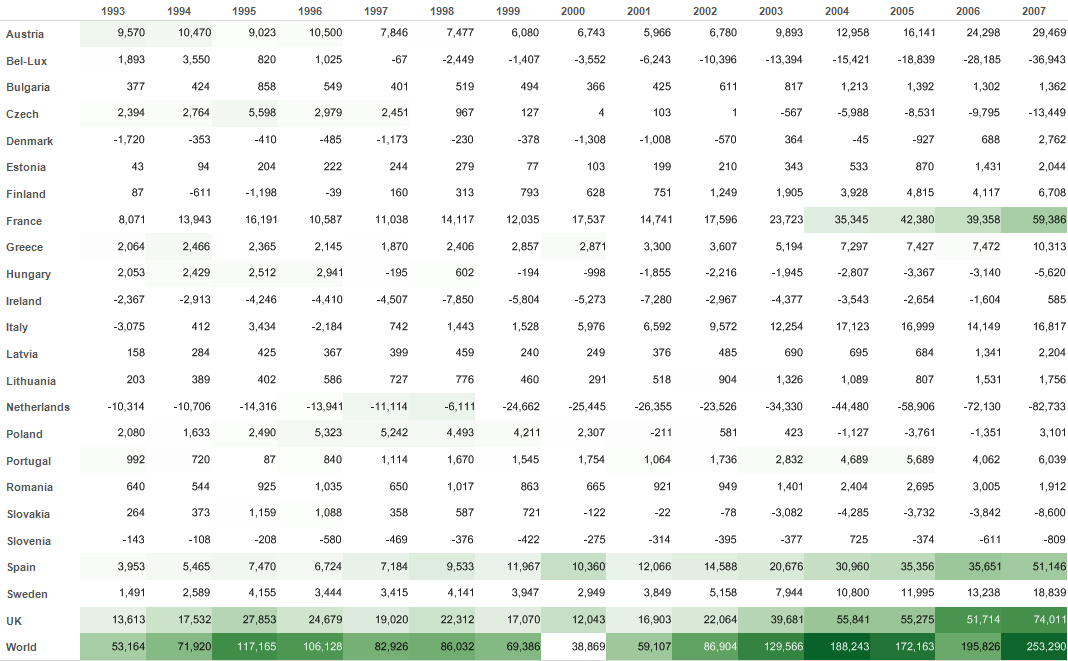}
\end{center}
\caption{ \trackchange{{\bf Evolution of the direct and indirect measures of trade imbalances for Germany.} The figures in each cell correspond to \trackchange{direct} trade surpluses (+) or deficits (-) of Germany toward countries listed on the rows. The colors correspond to the indirect measures of trade imbalances, as computed by the Flow Decomposition Method, with ultimate surpluses in green and ultimate deficits in red.}}
\label{colorsandnumbers_GER}
\end{figure}

\trackchange{More generally, the rising complexity and imbalances of the trade network, where some countries record large surpluses (China, Germany) and others large deficits (US, France) awake the debate on the need of a global coordinated adjustment plan involving all countries (see \cite{FS2009} and \cite{DF2013} for two recent network-related contributions on the global trade imbalances).  By shrinking the size of the trade information matrix, the indirect measure can help each country to identify the trade partners with which negotiating prioritarily an adjustment plan to correct the imbalances.} \trackchange{Fig. \ref{colorsandnumbers_GER}  refers to Germany and shows that trade deficits of many European countries toward Germany are netted out by network effect. Since France, Spain and UK have an ultimate deficit towards Germany, they are the prioritary countries with which the adjustment plan should be discussed.} 

\trackchange{Finally, the fact that Figs \ref{fig:flow}, \ref{colorsandnumbers_NED}, \ref{colorsandnumbers_FRA} and \ref{colorsandnumbers_GER}  exhibit quite regular, non-noisy trends further validates the method.}

\subsection*{Trade \trackchange{integration} with commute-time distance}

One objective of the Euro zone was to support the trade integration of its members. The trade integration (through lower transaction costs) is one of the main benefits derived from the common currency. We here illustrate how the network tools provided by the commute-time distance approach help to deepen the understanding of the European trade network, the \trackchange{degree of integration} of the countries in the network and \trackchange{their integration dynamics}. 

These distances, built upon the information available in the whole \trackchange{European} network, are global, \trackchange{or indirect}, measures of the country trade \trackchange{integration}, as opposed to the \trackchange{direct trade integration} computed as bilateral trade-to-GDP ratio. 

\begin{figure}[!ht]
\begin{center}
 \includegraphics[trim = 25mm 0mm 20mm 0mm,clip,width=\textwidth]{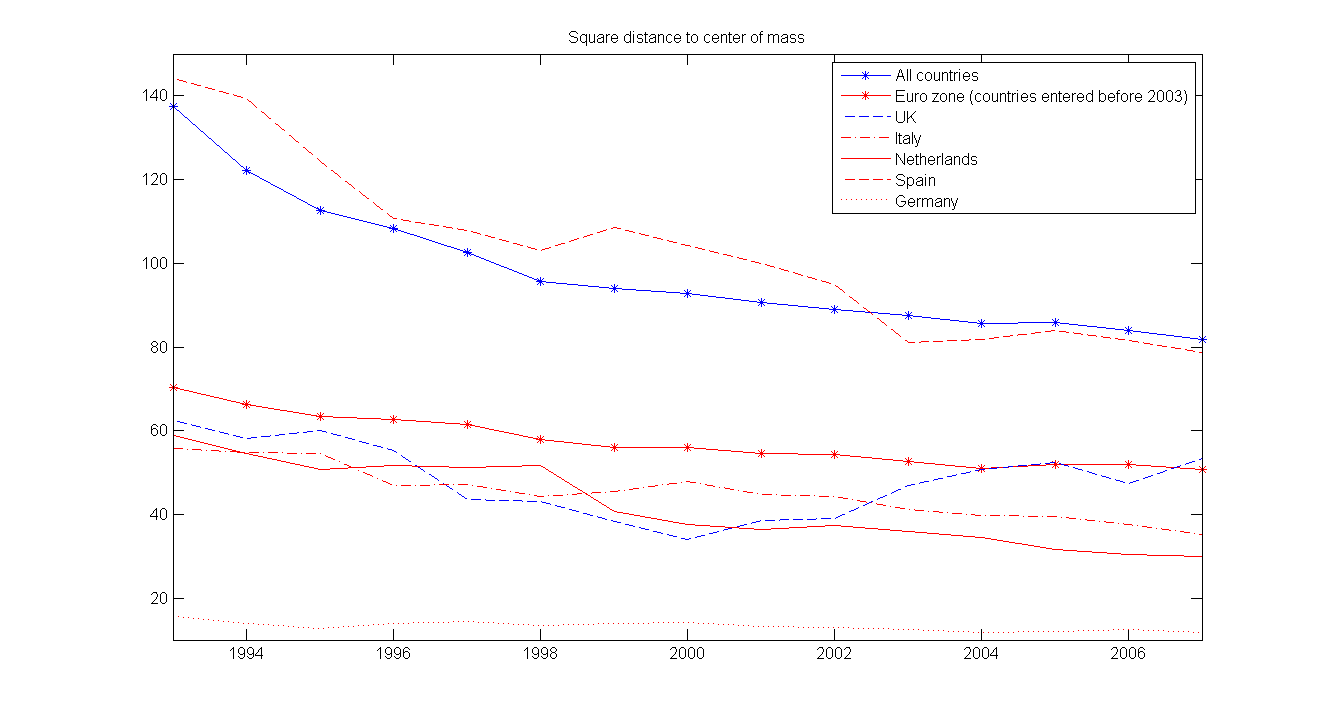}
\end{center}
\caption{
 \trackchange{{\bf Evolution of the variance of EU countries, Euro zone countries, and evolution of the square commute-time Euclidean distance of some countries to the centre of mass of Europe.} Centres of masses and variances are computed with weights proportionally to countries' PageRank. A falling distance to the centre of mass indicates an increasing integration of country's economy with its European partners'. A falling variance of a group of countries indicates a closer and closer integration of all countries with one another.}} 
\label{fig:fullVar}
\end{figure}

We first present in Fig. \ref{fig:fullVar} the variance of the \trackchange{distances of the Euro zone countries} \trackchange{to their centre of mass} \trackchange{over the period 1993-2007} on the one hand and of the 24 European countries of our sample on the other. The variance is computed as the sum of square distances to the centre of mass of the considered countries, \trackchange{each} weighted proportionally \trackchange{to} their PageRank. The variance allows to estimate with just one number how far away from each other the economies of the considered countries are. \trackchange{The larger the variance, the lower the mutual integration between the considered countries.} The PageRank weighting is meant to tone down the effect of small-size outliers such as Bulgaria in early nineties, \trackchange{and give more importance to large economies such as France or Germany. It can be verified that the centres of mass of the Euro zone and the whole European Union are very close, and rather close to Germany.}

\begin{figure}[!ht]
\begin{center}
 \includegraphics[trim = 25mm 0mm 20mm 0mm,clip,width=\textwidth]{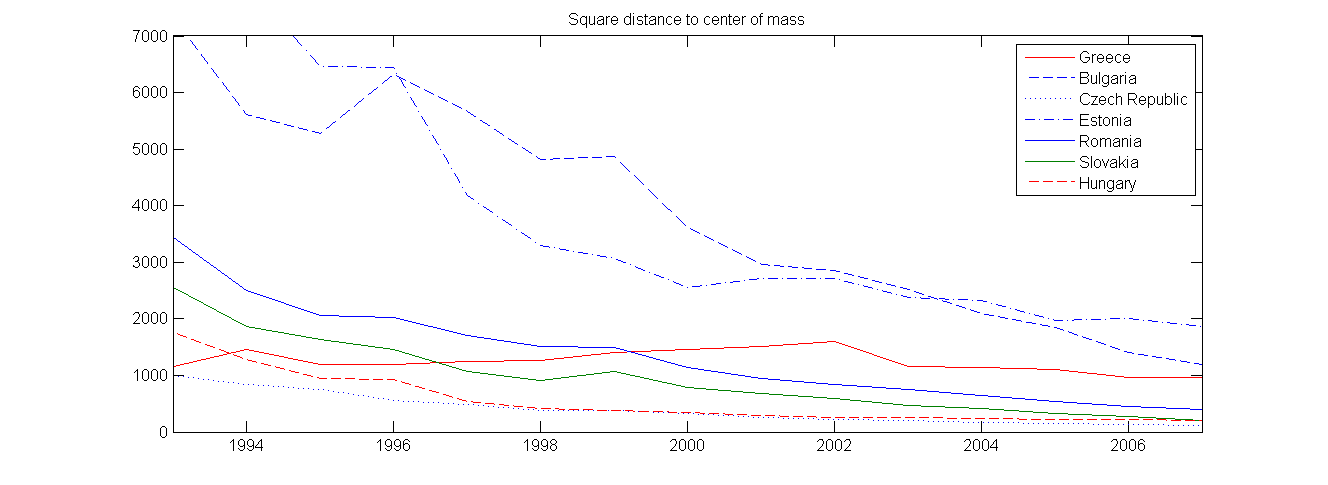}
\end{center}
\caption{
 \trackchange{{\bf Evolution of the square commute-time Euclidean distance of outlier economies to the centre of mass.} Note the contrast between the former socialist economies and Greece.}}
\label{fig:faraway}
\end{figure}

\trackchange{In the same figure, we also represent individual trajectories through their square distance to the centre of mass of the European Union, in order to show the variety of situations aggregated into the variance. For example, the United Kingdom, albeit outside the Euro zone, is actually better integrated to Euro zone than some Euro zone countries such as Spain. Nevertheless from the years 2000 it tends to split away from the centre of mass.  Interestingly, this tendency is also observed for some Euro zone countries as we observe in another visualization hereafter. Fig. \ref{fig:faraway} further illustrates the diversity of situations, focusing on the small European economies lying far away from the centre of mass. We can see that the Eastern Europe \trackchange{economies, while starting very far away from the centre in 1993, steadily close the gap, while for instance Greece remains at constant distance and even drifts away for a few years.} \trackchange{From a methodological perspective, we also see that each curve exhibits clear and consistent trends, showing that the tool delivers non-noisy results.}}

The comparison reveals that the average inter-country trade \trackchange{distances}, as measured by their variance, is much smaller for Euro zone countries than for European countries (in or out of the Euro zone), revealing a larger integration among Euro countries. We also note that the variance decreases, mostly at the beginning of the period, with no specific break in 1999 for the introduction of the Euro. Some preliminary studies estimated the effect of the launch of the Euro on the volume of trade within the Euro zone countries to be in the range of 8\% to 16\% (much less than the 200\% effect detected by Andrew Rose in his seminal paper on the currency unions \cite{Baldwin2006, Engel2004, Rose2000}\trackchange{)}. That the graphic of the variances does not capture any substantial effect related to the Euro in 1999 is only half a surprise to the extent that most of the integration is known to have occurred before, especially with the Single Market Program of 1992. In addition, the absence of effect in our \trackchange{integration} measure could also reveal some centrifugal forces for countries whose trade mostly occurs with countries at the periphery of the Euro zone. This is confirmed below in the country specific analysis.

\begin{figure}[!ht]
\begin{center}
 \includegraphics[width=14cm]{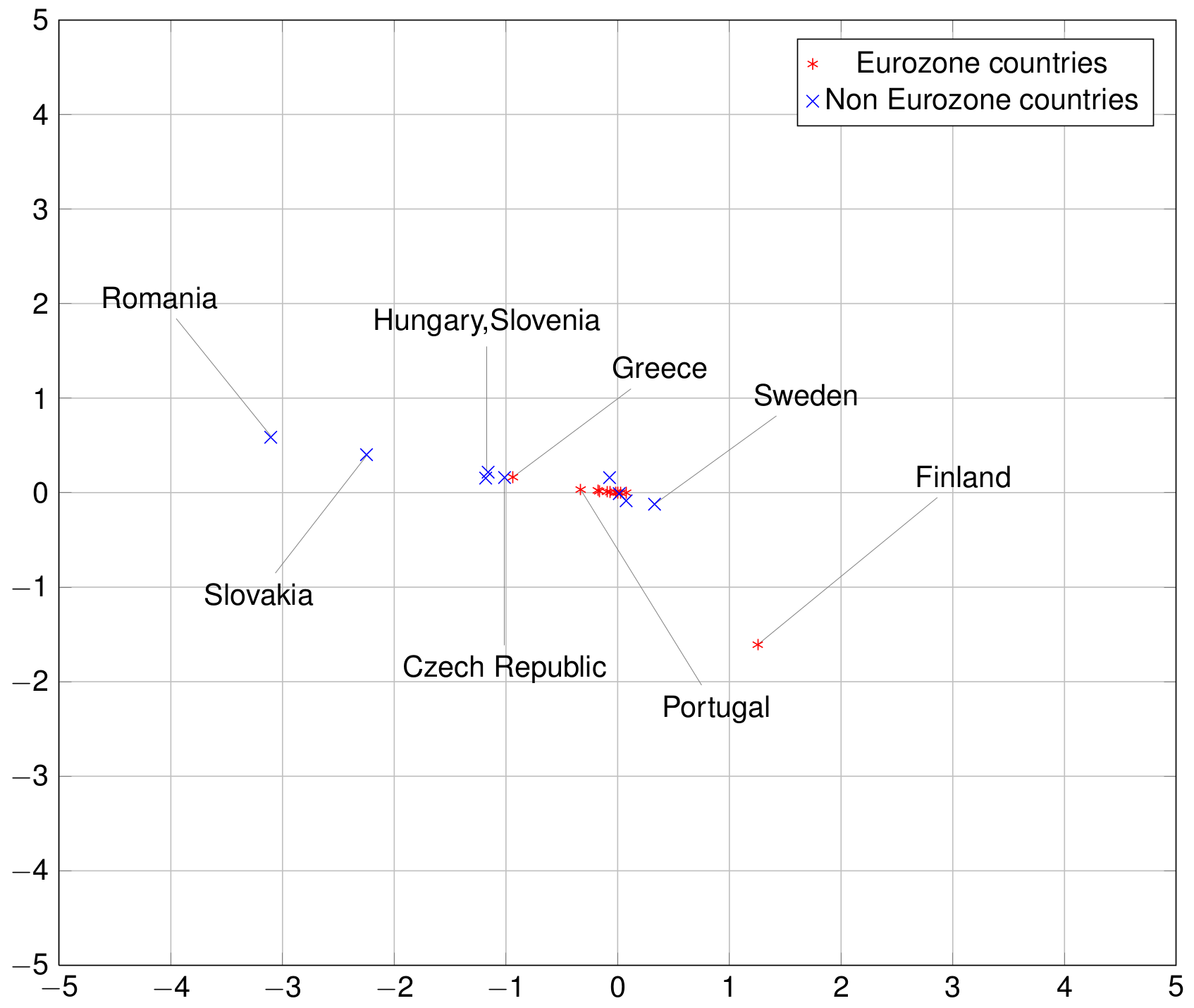}
\end{center}
\caption{
{\bf European economies in 1993.} The square Euclidean distance between two countries \trackchange{in this 2D representation} is an approximation of the random dollar's commute time between those two countries. Compare with Figs \ref{Dispersions2000} and \ref{Dispersions2007}. Note that some countries are outside the bounding box, especially some East European countries such as Bulgaria in the early years. Among Euro zone countries, notice how Greece, Finland and Portugal stay relatively \trackchange{far} from the Euro zone cluster and even drifts away from it after the introduction of the Euro.}
\label{Dispersions1993}
\end{figure}

\begin{figure}[!ht]
\begin{center}
 \includegraphics[width=14cm]{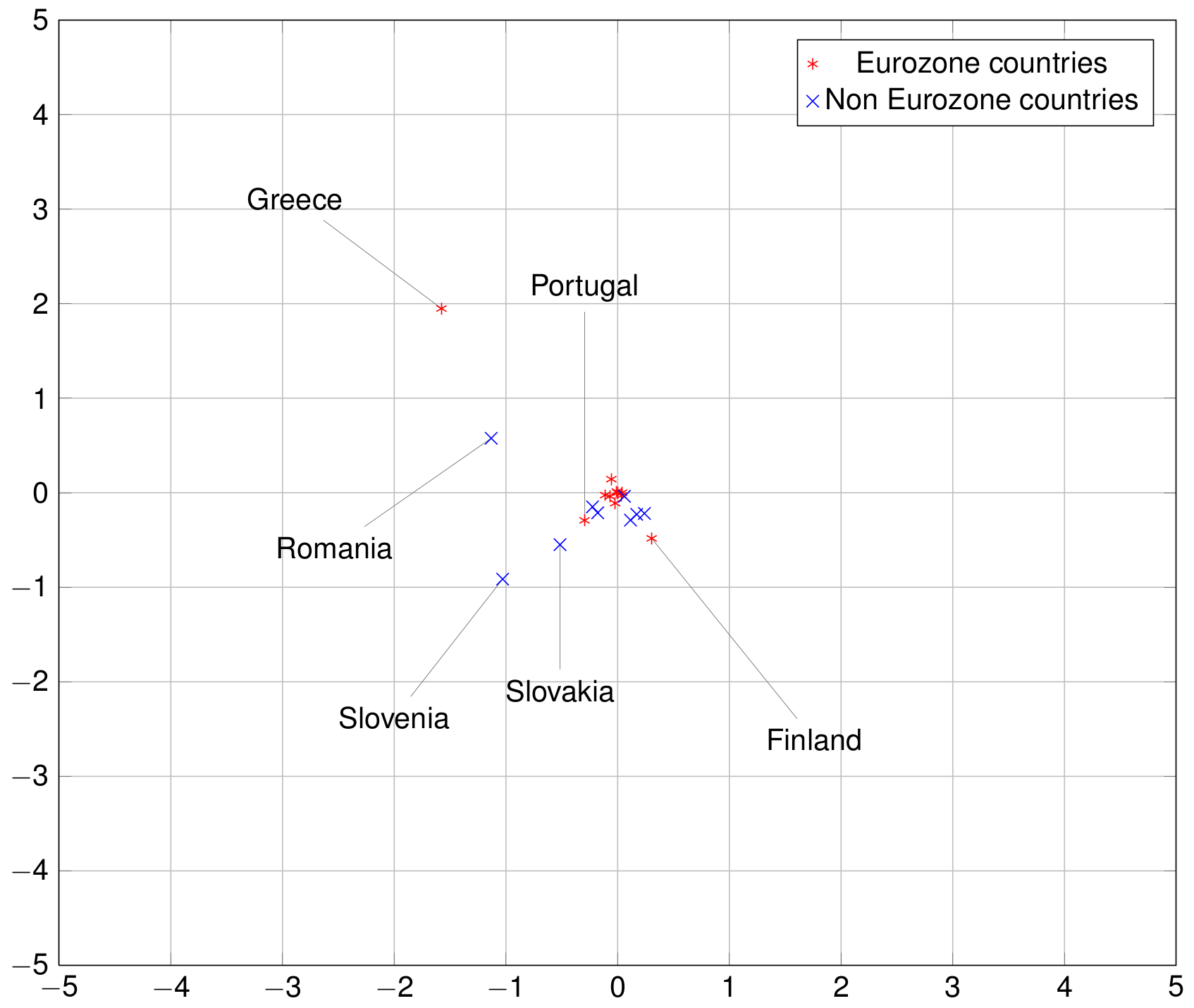}
\end{center}
\caption{
{\bf European economies in  2000.}  \trackchange{Compare with Figs \ref{Dispersions1993}} and \ref{Dispersions2007}.
}
\label{Dispersions2000}
\end{figure}

\begin{figure}[!ht]
\begin{center}
 \includegraphics[width=14cm]{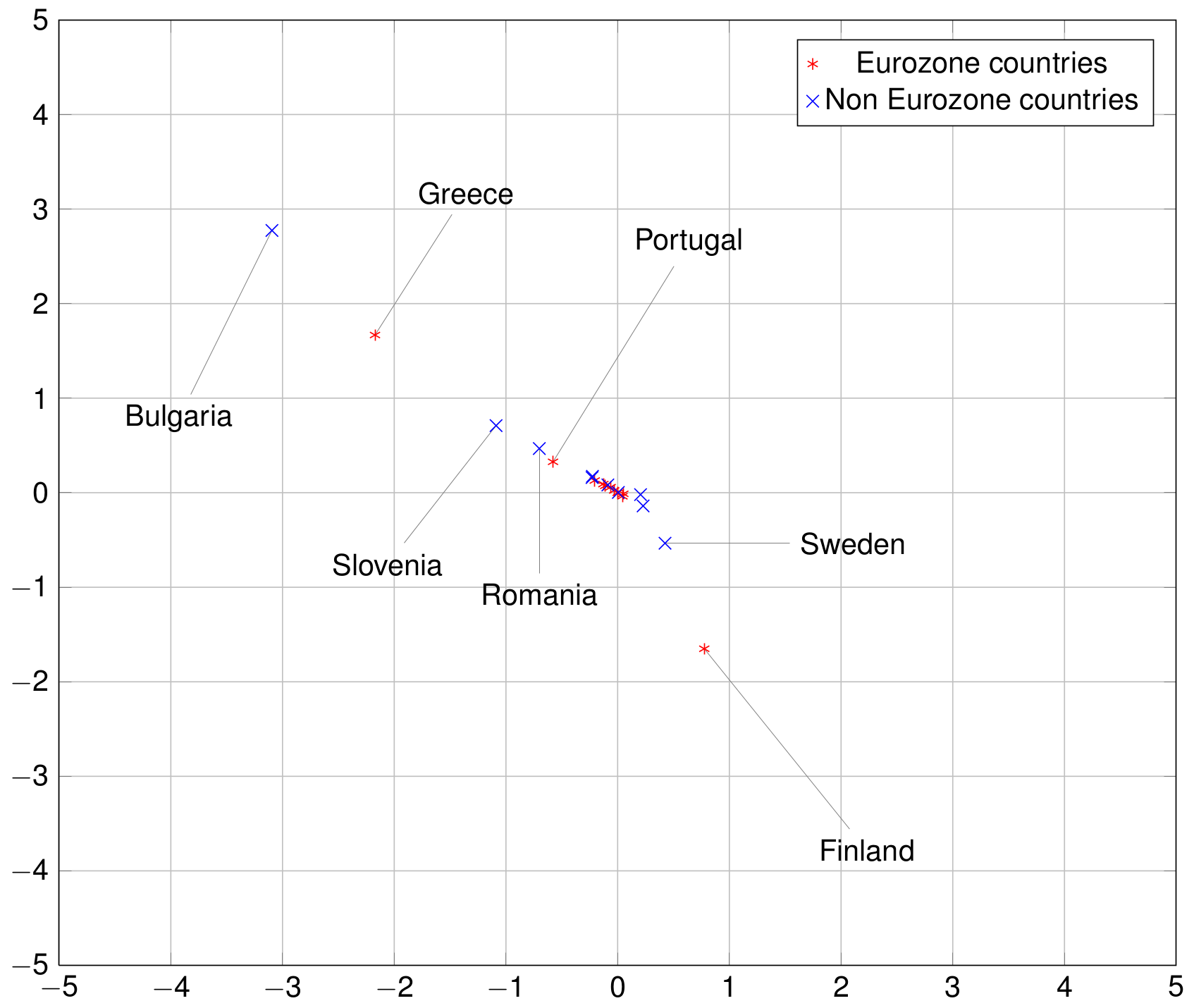}
\end{center}
\caption{
{\bf European economies in  2007.}  \trackchange{Compare with Figs \ref{Dispersions1993} and \ref{Dispersions2000}}.
}
\label{Dispersions2007}
\end{figure}

\trackchange{The commute-time distances of the 24 countries considered can be represented exactly in a 24-dimensional embedding of the national economies.}
\trackchange{Given the difficulty of a  24-dimensional visualization, we project this representation  in a two-dimensional space by principal component analysis}.
\trackchange{These pictorial representations of European economies} in Figs \ref{Dispersions1993} to \ref{Dispersions2007}, reveal that some \trackchange{Euro zone} countries started in about 2000 to take distance from the core of the network. This evolution was not necessarily captured by \trackchange{direct trade integration measures such as trade-to-GDP ratio}. This is mostly due to countries like Finland, Greece and, to a lesser extent, Portugal. As regards Finland, the \trackchange{increasing distance to the core} is mainly due to two factors. First, the share of the trade with the European countries decreased particularly over the period 2002-2007, as captured by the \trackchange{trade-to-GDP ratio}. Second, the share of exports to large European countries such as Germany or France also decreased, which is not captured by the \trackchange{trade-to-GDP ratio}. The combination of those factors explain the relative divergence of Finland. On the contrary, for Greece, the divergence in the trade network \trackchange{integration} two-dimensional representation is not due to a decreasing share of trade with the European Union, but to a large decrease of trade with core countries (such as Germany). As regards Portugal, the share of trade with the European countries remains globally stable over the 1993-2007 period, but an acuter integration with Spain, raised the relative share of this peripheral countries and reduced the share of other more central countries.

%\clearpage
%\newpage

%\clearpage
%\newpage

%The commute-time trade \trackchange{integration} measure thus complements the direct trade-to-GDP integration measure. \trackchange{For} Portugal and Greece, the \trackchange{latter} measure did not detect the shift of trade composition, contrarily to the \trackchange{commute-time} measure which captured the decreasing share of trade with countries in the core of Europe. 

\section*{Conclusion}

%\trackchange{[JC:  Future work: transmission of economic shock; dialectics central periphery]}

As suggested by Fracasso and Schiavo \cite{FS2009} in their network analysis of global imbalances and by Schweitzer et al. \cite{Science} in their review of economic network challenges, measuring economic variables and accounting for the network complexity requires new tools and new approaches. The Flow Decomposition Method, as a network counterpart for the trade deficit-to-GDP ratio, and the Commute Time Distance method as a network counterpart for the trade-to-GDP ratio, are shown to shed more light on the evolution of EU economies before and after the introduction of a common currency. 

The launch of the Euro and the recent threats to its perennity surely stress the importance of better apprehending the European trade structure. \trackchange{The trade integration indirect measure helps to better identify those countries sliding out of the trade union. The trade imbalances indirect measure helps to better apprehend who are the ultimate debtors and creditors in the trade network.}

The techniques used here are of course of interest for general networks, and while the Commute Time Distance methods and related graphical representation are standard tools reinterpreted here through their economic meaning, the Flow Decomposition Method introduced in this paper can help to analyse any kind of flows on graphs, e.g. car traffic on a road network.

% Do NOT remove this, even if you are not including acknowledgments
\section*{Acknowledgments}

G.K. and J.-C. D. are supported by the Belgian Programme of Interuniversity
Attraction Poles initiated by the Belgian Federal Science Policy Office and an Action de Recherche Concertée (ARC) of the French Community of Belgium. 
\trackchange{Conversations with Florian Mayneris helped improve the paper thoroughly.}

%\section*{References}
% The bibtex filename
\bibliography{citations}

%\clearpage
%\newpage
%\section*{Figure Legends}
%\begin{figure}[!ht]
%\begin{center}
%%\includegraphics[width=4in]{figure_name.2.eps}
%\end{center}
%\caption{
%{\bf Bold the first sentence.}  Rest of figure 2  caption.  Caption 
%should be left justified, as specified by the options to the caption 
%package.
%}
%\label{Figure_label}
%\end{figure}

%\clearpage
%\newpage

%\clearpage
%\newpage

%\clearpage
%\newpage
%\section*{Tables}
%\begin{table}[!ht]
%\caption{
%\bf{Table title}}
%\begin{tabular}{|c|c|c|}
%table information
%\end{tabular}
%\begin{flushleft}Table caption
%\end{flushleft}
%\label{tab:label}
% \end{table}

\end{document}